\documentclass{ecai}
\usepackage{graphicx}
\usepackage{latexsym}
\usepackage{times}
\usepackage{soul}
\usepackage{url}
\usepackage[hidelinks]{hyperref}
\usepackage[utf8]{inputenc}
\usepackage[small]{caption}
\usepackage{graphicx}
\usepackage{amsmath}
\usepackage{amsthm}
\usepackage{booktabs}
\usepackage{algorithm}
\usepackage{algorithmic}
\usepackage{multirow}

\begin{document}

\begin{frontmatter}

\title{TRLS: A Time Series Representation Learning Framework via Spectrogram for Medical Signal Processing}

\author[A]{\fnms{Luyuan}~\snm{Xie}\orcid{0000-0002-4777-122X}\thanks{This work was supported by the National Key R$\&$D Program of China under Grant No.2022YFB2703301.}}
\author[A]{\fnms{Cong}~\snm{Li}}
\author[A]{\fnms{Xin}~\snm{Zhang}} 
\author[A]{\fnms{Shengfang}~\snm{Zhai}} 
\author[A]{\fnms{Yuejian}~\snm{Fang}} 
\author[A]{\fnms{Qingni}~\snm{Shen}} 
\author[A]{\fnms{Zhonghai}~\snm{Wu}}

\address[A]{School of Software and Microelectronics, Peking University, Beijing, China}
\begin{abstract}
Nowadays, a series of representation learning frameworks in unlabeled time series have been proposed, which have numerous practical applications. However, these frameworks can not generate sufficiently robust representations for downstream tasks. The main reasons for this phenomenon are that they suffer from three major flaws: (1) The time domain of time series is usually complex dynamic and annotated sparsely, and the current frameworks don't well overcome these difficulties; (2) These frameworks often face the problem of constructing fragile negative samples; (3) They cannot provide the robust enough encoder. Consequently, building the representation learning framework in unlabeled time series is still a challenging task. In this paper, we present a \textbf{T}ime series \textbf{R}epresentation  \textbf{L}earning framework via \textbf{S}pectrogram (TRLS) to learn more robust representations. In our TRLS, as an input, the spectrogram contains time and frequency domain information which can alleviate the complexity and sparsity of time series. In the training stage, we maximize the similarity between positive ones, which effectively circumvents the problem of designing negative samples. Finally a novel encoder named Time Frequency RNN (TFRNN) is designed to extract more robust multi-scale representations from the augmented spectrogram. Our evaluation on four real-world time series datasets shows that TRLS is superior to other existing frameworks.  We will open-source our code when the paper is accepted.
\end{abstract}

\end{frontmatter}

\section{Introduction}
Time series plays a crucial role in various  fields, including medical and financial domains, such as ECG prediction and stock analysis \cite{oreshkin2019n}. With the development of IoT and wearable devices, it is more convenient to collect time series \cite{eldele2021time}. However, the annotation of time series is greatly limited due to the high requirement for professional knowledge in time series analysis \cite{ching2018opportunities}. Additionally, deep learning typically requires a large amount of labeled data for training. These issues make learning representations from unlabeled time series a significant and meaningful challenge.


\begin{figure}[htb]
\setlength{\belowcaptionskip}{-0.4cm}
\centering
\includegraphics[width=.5\textwidth]{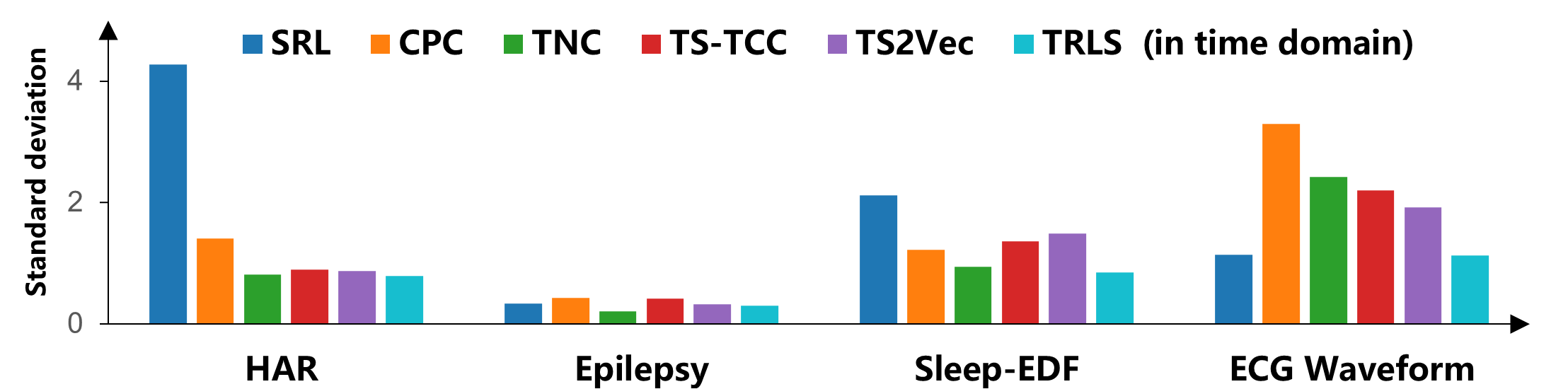}
\caption{The accuracy standard deviation of the time series contrastive learning frameworks under 5-fold cross validation for four datasets.}
\label{fig:traditional}
\end{figure}
In recent years, self-supervised learning has attracted wide attention. It extracts effective data representations from unlabeled data for downstream tasks \cite{chen2020simple,cheng2020subject}. As a kind of self-supervised representation learning, contrastive learning \cite{DBLP:conf/icml/ChenK0H20} performs well in many downstream tasks. Therefore, numerous self-supervised representation learning and contrastive learning frameworks \cite{ching2018opportunities,franceschi2019unsupervised,tonekaboni2021unsupervised,eldele2021time} have been proposed in the time series analysis field. The above frameworks have been proven to achieve superior performance. However, due to the inability to generate sufficiently robust representations, these frameworks may have significant differences in accuracy when trained on different subsets of the same dataset. In this paper, robust refers to the ability of the model to achieve good and stable performance across different tasks or datasets. As shown in Figure 1,  we conduct 5-fold cross-validation experiments on four public datasets under the same experimental conditions (i.e., the encoder and the subsets of datasets are consistent) for these frameworks (SRL, CPC, TNC, TS-TCC and TS2Vec). They have a large standard deviation in accuracy. The main reasons for this phenomenon are that they suffer from three major flaws. First, owing to only taking time domain signals of time series with complex dynamics and sparse annotations as inputs, it is difficult for these frameworks to generate enough appropriate representations. Then some studies have shown that constructing fragile negative samples may reduce the effectiveness of these self-supervised representation learning frameworks \cite{grill2020bootstrap,chen2020simple}. Finally, these frameworks merely use TCN or RNN as the encoder \cite{tonekaboni2021unsupervised} which can not extract the multi-scale representations for more robust representations.

In this paper, we propose a  \textbf{T}ime series \textbf{R}epresentation  \textbf{L}earning framework via \textbf{S}pectrogram (TRLS) based on time-frequency analysis. Firstly, to tackle the problem of time series with complex dynamics and sparse annotations, we convert the time series into a spectrogram through short-time Fourier transform (STFT) \cite{jeon2020area}. The spectrogram not only retains time domain information of time series, but also aggregates differentiated frequency domain information. Moreover, the flexible window size can handle time series of varying lengths. To augment the data, image data augmentations are adapted for spectrogram as existing time series data augmentation methods mainly act on the time domain. TRLS can train more robust encoder by maximizing the similarity between positive samples in the training stage and avoiding the problem of constructing fragile negative samples. Regarding the encoder, we design a new encoder called Time-Frequency RNN (TFRNN) to extract multi-scale representations of the augmented spectrogram. In summary, the main contributions of this work are as follows.

\begin{itemize}
    \item We propose TRLS, a novel time series representation learning framework which uses spectrograms obtained through STFT to generate more robust representations. TRLS addresses the challenges of complex dynamics and sparse annotations in time domain signals, and applies image data augmentations to the spectrogram.
    \item Our framework uses only positive samples for encoder training, effectively solving the problem of negative sample design in the training stage.
    \item TRLS incorporates a new encoder, TFRNN, that extracts multi-scale representations of the augmented spectrogram to learn more robust representations of time and frequency.
    \item We evaluate TRLS on four public datasets and show that it outperforms current state-of-the-art frameworks.
\end{itemize}

\begin{figure*}[htb]
\setlength{\belowcaptionskip}{-0.4cm}
\centering
\includegraphics[width=.9\textwidth]{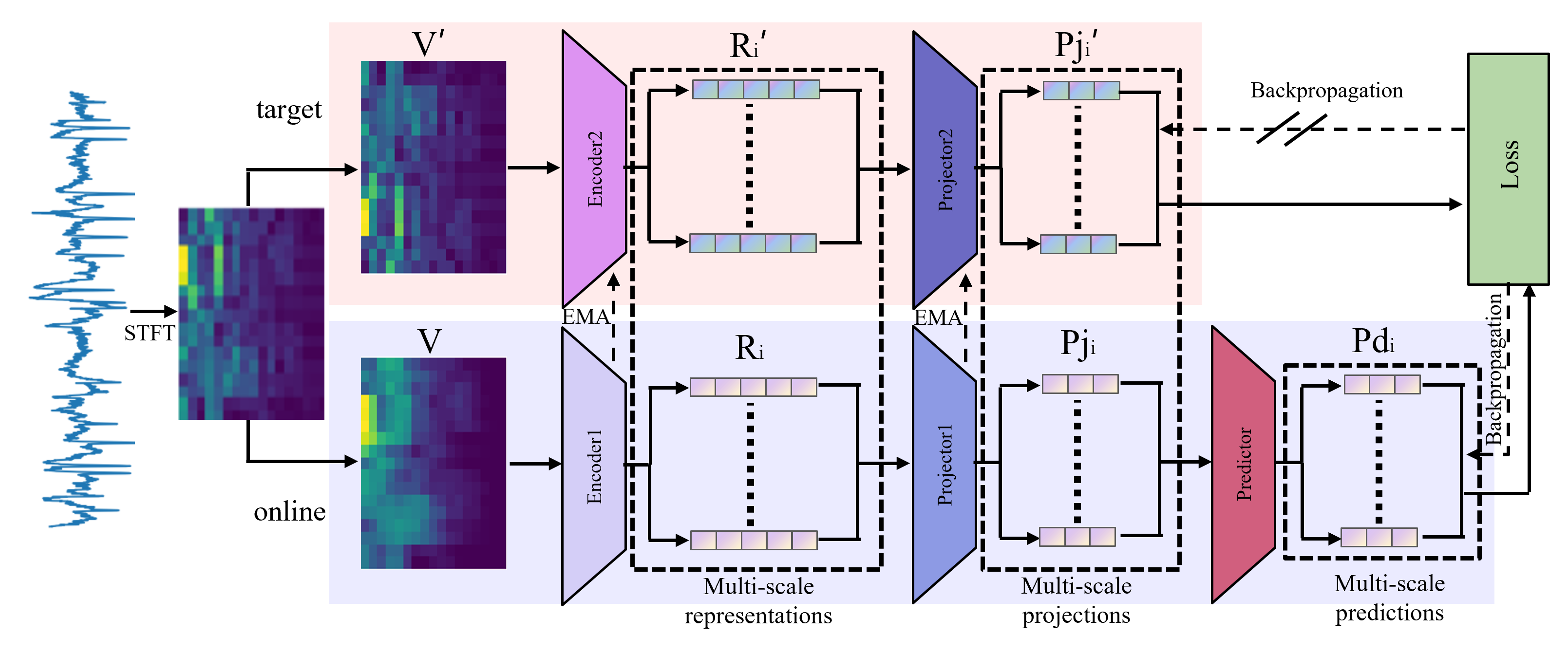}
\caption{ Overall architecture of proposed TRLS framework.}
\label{fig:traditional}
\end{figure*}

\section{Related Works}
\subsection{Self-supervised Learning}


Self-supervised learning has been popularized in computer vision for learning representations through pretext tasks, such as solving jigsaw puzzles \cite{noroozi2016unsupervised}, image colorization \cite{zhang2016colorful}, and predicting image rotation \cite{gidaris2018unsupervised}. While these tasks have shown excellent performance, their reliance on heuristics may limit the effectiveness of the learned representations. Contrastive learning, which focuses on augmenting invariance of data representation, has recently gained attention. MoCo v1 \cite{he2020momentum} proposes a Momentum encoder to learn representations of negative pairs obtained from a memory bank, while SimCLR \cite{DBLP:conf/icml/ChenK0H20} uses a larger batch of positive and negative pairs without a momentum encoder. BYOL \cite{grill2020bootstrap} uses bootstrapping representations to learn features without negative samples, and SimSiam \cite{chen2021exploring} utilizes a Siamese network and stop gradient to improve the effect of representation learning. Some improvements have also been made based on the encoder, such as MoCo v3 \cite{chen2021empirical} replacing the CNN encoder with ViT and CARE \cite{ge2021revitalizing} introducing the transformer to guide the training of CNN encoder. However, these schemes may not be directly applicable to time series due to their different attributes.

\subsection{Self-supervised Learning for Time Series}

At present, a large number of representation learning for time series methods have been proposed. SSL-ECG \cite{sarkar2020self} learns the representations of ECG by using six methods of data conversion and assigning pseudo tags to datasets. \cite{aggarwal2019adversarial} learns subject invariant representations through local and global modeling of data. BTSF \cite{yang2022unsupervised} leverage time-frequency pairs to learn representations from time series. TST \cite{2021A} combines transformer to achieve multivariate time series representation learning.  Contrastive learning has shown to be a promising approach for representation learning in time series, and many recent studies in this area have adopted contrastive learning as the main approach. SRL \cite{franceschi2019unsupervised} proposes a novel triple loss method combined with random sampling to learn scalable representation. CPC \cite{oord2018representation} uses the autoregressive model and noise contrast method for representation learning. TNC \cite{tonekaboni2021unsupervised} proposes a neighborhood-based unsupervised learning framework for non-stationary time series. TS-TCC \cite{eldele2021time} introduces strength augmentations and disturbances with different time stamps to learn robust representation. TS2Vec \cite{yue2021ts} applies hierarchical contrasting to learn scale-invariant representations within augmented context views. The main difference between the five contrastive frameworks is that they select contrastive pairs according to different segment-level sampling policies. They have achieved impressive results. However, the sparsity and complexity of time series make them difficult to learn more robust representation. In addition, designing more robust encoders and the problem of constructing negative samples also affects the performance of representation learning. In this paper, we propose a new framework to solve the above problems.

\section{Methods}
In this section, we provide the detailed description of the TRLS framework. Section 3.1 introduces the TRLS framework. Section 3.2 presents the concept of spectrograms and the data augmentation. Section 3.3 provides a detailed explanation of the training process employed for TRLS, while Section 3.4 discusses the encoder TFRNN.


\begin{figure}[htb]
\setlength{\belowcaptionskip}{-0.4cm}
\centering
\includegraphics[width=.48\textwidth]{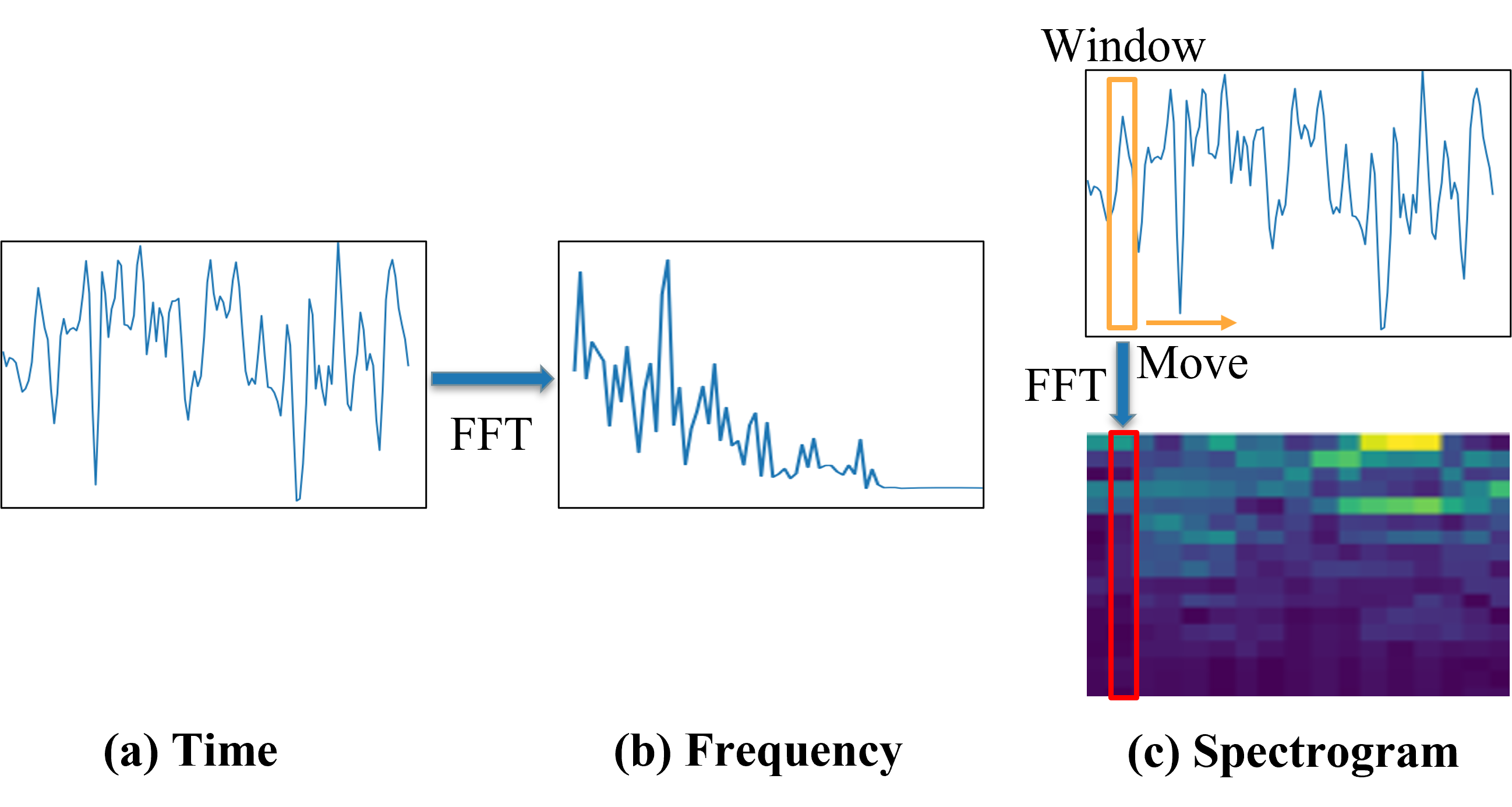}
\caption{The different inputs of the time series representation frameworks.}
\label{fig:traditional}
\end{figure}

\subsection{TRLS Framework}

The aim of our proposed TRLS is to learn more effective representations $R_i$ ($i$ $\in$ $K$, $K$ denotes the number of representations) for downstream tasks. As depicted in Figure 2, our TRLS comprises two networks: the online network and the target network. The online network consists of three components: an encoder, a projector, and a predictor. The target network has the same encoder and projector architecture as the online network but uses a distinct set of weights.
Initially, we perform STFT operations on the input data to obtain a spectrogram, to which we apply two different data augmentations. Next, we use an encoder called TFRNN to extract multi-scale representations of the augmented spectrogram, followed by projectors to reduce the dimensions of these multi-scale representations. In contrast to the target network, the online network includes a predictor that improves the encoder's representation capability. During training, we use the multi-scale projections from the target network and the multi-scale predictions generated by the online network for loss calculation. Additionally, the online network is updated by the gradient, while the target network is updated according to the Exponential Moving Average (EMA) \cite{grill2020bootstrap} with the online network. 

\subsection{Spectrogram and Data Augmentation}
One of difference between TRLS and mainstream frameworks which operate on time domain data (Figure 3 (a)) \cite{ching2018opportunities,franceschi2019unsupervised,tonekaboni2021unsupervised,eldele2021time} or separate time and frequency domains (Figure 3 (a) and (b)) \cite{yang2022unsupervised,zhang2022tfad} as inputs is that TRLS uses STFT to obtain the spectrogram (Figure 3 (c)). STFT is a method that decomposes time series into time-frequency domain information by dividing the long-time signal into short-time segments by moving window and performing Fast Fourier Transform (FFT) on each segment to obtain the frequency spectrum information for that segment. This makes the spectrogram obtained through STFT preserve both temporal and spectral information. 
Compared to FFT, which only provides frequency-domain information for the entire signal, STFT has better time-frequency localization characteristics and can capture the instantaneous changes of the signal in time and frequency \cite{jeon2020area,yang2022unsupervised,zhang2022tfad}. As a result, features generated by STFT are more effective than FFT or time domain features in some tasks. Furthermore, STFT can adjust the time-frequency resolution and smoothness by selecting different window sizes and overlap lengths to suit time series data of varying lengths. 

Data augmentation is a crucial component of contrastive learning methods \cite{chen2020simple,cheng2020subject} and has been shown to significantly influence the performance of trained encoders in recent research \cite{eldele2021time,yang2022unsupervised,chen2020simple}. It is therefore imperative to design appropriate data augmentation strategies for our proposed framework. Traditional contrastive learning methods often use two (random) variants of the same augmentation, which can negatively impact the robustness of learned representations. To avoid this issue, we follow the approach in \cite{eldele2021time} and employ two different data augmentation methods. While most time series contrastive learning methods rely solely on time-domain data augmentation, we also incorporate image augmentation techniques on the spectrogram \cite{xie2022tc}, such as ColorJitter and GaussianBlur \cite{ranjan2017optical}. Our experimental results in Section 5 indicate that direct spectrogram augmentation is optimal. Specifically, in TRLS, we apply ColorJitter and RandomHorizontalFlip for the first augmentations and Gaussian Blur and RandomHorizontalFlip for the second augmentations.

\subsection{Training without Negative Samples}

The proposed TRLS framwork operates on time series $T$ by using STFT and two different data augmentations to generate the spectrogram in two different augmented views, denoted as $V$ and $V'$. The online network takes the first augmented view $V$ as input and produces multi-scale representations $R_i$, as well as multi-scale projections $Pj_i$ and $Pd_i$. In contrast, the target network uses the second augmented view $V'$ and outputs $R_i'$ and multi-scale projections $Pj_i'$. Notably, the predictor is only applied to the online network, resulting in an asymmetric architecture between the online and target pipelines. Additionally, a mean squared error (MSE) is defined to evaluate the normalized multi-scale predictions against the target multi-scale projections.

\begin{align}
            \resizebox{.91\linewidth}{!}{$
            L = \frac{1}{K}\sum\limits_{i = 1}^K {\left\| {{{\overline {Pd} }_i} - \overline {Pj} _i'} \right\|} _2^2{\rm{ }} = \frac{1}{K}\sum\limits_{i = 1}^K {2{\rm{ }} - {\rm{ }}2 \times \frac{{\left\langle {P{d_i},Pj_i'} \right\rangle }}{{{{\left\| {P{d_i}} \right\|}_2} \times {{\left\| {Pj_i'} \right\|}_2}}}}
            $}
\end{align}%
We symmetrize the loss function $L$ in Eq.1 by separately feeding {$V{'}$} to the online network and $V$ to the target network to compute $L{'}$, respectively.
\begin{align}
            {L_{total}}{{  =  L  +  L'}}
\end{align}%
$L_{total}$ is used to maximize the similarity between positive samples for training the encoder. This means that the TRLS don't require negative samples in the training stage, which effectively avoids the problems caused by the inappropriate negative samples. By performing an adam optimization step, we minimize $L_{total}$ for updating the online network’s parameters $\beta$ at each training step. The target network is depicted by the stop gradient in Figure 2. Target network’s parameters $\delta$ update by:
\begin{align}
            \delta  \leftarrow \tau \delta  + (1 - \tau )\beta 
\end{align}%
$\tau$ is the moving average decay.


\subsection{The encoder for the spetcrogram}
Temporal CNN (TCN) \cite{bai2018empirical} is used in the time series contrastive learning frameworks frequently. It adopts 1D convolution that can well extract time series time domain features. Actually, apart from time series domain information, spectrogram also includes frequency domain information, which results in that TCN may lose frequency information when extracting time dimension information of spectrogram. To cope with this problem, we design a Time and Frequency RNN (TFRNN) to learn the time-frequency characteristics of the spectrogram. As shown in Figure 4 (a), the Time-Frequency RNN block applies different RNN for the time and frequency of the spectrogram accordingly.

For any given feature map $X \in t$ × $f$ ($t$ represents the time dimension, and $f$ represents the frequency dimension), it first generates feature $Y$ through Time RNN (TRNN):

\begin{align}
    {{Y  =  {\rm TRNN}(X) \quad Y}} \in {{t}} \times {{h}}
\end{align}%
where h represents the hidden layer node of TRNN. Then $Y$ goes through Feed Forward layer (FF) to keep consistent with the frequency dimension of input feature $X$:
\begin{align}
            {{{Y}}_{{1}}}{{\rm = FF}(Y)} \quad {{{Y}}_{{1}}} \in {{t}} \times {{f}}
\end{align}%
Through Batch Normalization (BN) and adding the input feature $X$ to get $Y_{t}$ $\in t$ × $f$:
\begin{align}
            {{{Y}}_{{t}}}{{\rm  =  BN(}}{{{Y}}_1}{{)  +  X    }}  \quad {{{Y}}_{{t}}} \in {{t}} \times {{f}}
\end{align}%
Before $Y_{t}$ $\in t$ × $f$ entering Frequency RNN (FRNN), transpose $Y_{t}$ to obtain $Y_{2}$ $\in f$ × $t$:
\begin{align}
            {{{Y}}_{{2}}}{{  =  Y}}_{{t}}^{{T}}\quad{{  Y}} \in f \times t
\end{align}%
The next operation is similar to formula (4) (5), which is expressed as:
\begin{align}
           {{{Y}}_3}{{\rm  =  FRNN(}}{{{Y}}_2}{{)}}\quad{{{Y}}_3} \in {{f}} \times {{{h}}_1}
\end{align}%
\begin{align}
            {{{Y}}_{{4}}}{{\rm=FF(}}{{{Y}}_3}{{)}}\quad{{{Y}}_4} \in {{f}} \times {{t}}
\end{align}%
where $h_{1}$ represents the hidden layer nodes of FRNN. $Y_{3}$ maintains consistency with the time dimension of input feature $X$ through Feed Forward (FF). Finally, $Y_{4}$ is added with $Y_{t}$ after transposing operation T and BN to get $Y_{out}$ $\in t$ × $f$.
\begin{align}
            {{{Y}}_{{{out}}}}{{\rm  =  BN}(Y}_4^{{T}}{{)  +  }}{{{Y}}_{{t}}}{{    }}\quad{{{Y}}_{{{out}}}} \in {{t}} \times {{f}}
\end{align}%

Based on TFRblock, we propose TFRNN (Figure 4 (c)) as the encoder of the TRLS. After the input spectrogram passes through three layers of TFRNN, the high-dimensional mapping $f$ $\rightarrow$ $f_{h}$ of the frequency domain dimension is performed with the 1D convolution (kernel size = 1) to obtain the feature $H \in t$ × $f_{h}$. Then, the pyramid pooling \cite{lin2017feature} is adopted and max pooling is used for multi-scale downsampling in the time dimension of feature $Y_{out}$ to get multi-scale features $D_{i}$:
\begin{align}
            {{{D}}_{{i}}}{{ = }}\left\{ {\begin{array}{*{20}{c}}
{{\rm{Maxpooling(}}{{{D}}_{{{i - 1}}}}{{) }}}\\
{{{{D}}_{{1}}}}
\end{array}\begin{array}{*{20}{c}}
{2 \le i \le {K}}\\
{i = 1}
\end{array}} \right.{\rm{  }}
\end{align}%
where K refers to the number of downsampling. The global average pooling (GAP) \cite{lin2013network} of the time dimension is performed on the multi-scale features $D_{i}$ to obtain multi-scale representations $R_{i}$.

\begin{table*}[htbp]
  \centering
      \caption{Comparisons between our proposed TRLS framework against baselines using linear classifier evaluation experiment.}
    \begin{tabular}{ccccccccc}
    \toprule
    \multirow{2}[4]{*} & \multicolumn{2}{c}{HAR} & \multicolumn{2}{c}{Epilepsy} & \multicolumn{2}{c}{Sleep-EDF} & \multicolumn{2}{c}{ECG Waveform} \\
          & ACC   & MF1   & ACC   & MF1   & ACC   & MF1   & ACC   & MF1 \\
    \midrule
    Supervised & 92.44±1.09 & 90.32±0.88 & 97.33±1.09 & 95.62±0.59 & \textbf{85.80±1.32} & 74.76±0.40 & 84.67±2.37 & 67.36±1.24 \\
    SRL   & 64.70±4.27 & 62.37±1.37 & 87.62±0.33 & 83.47±0.69 & 77.31±2.12 & 67.73±0.57 & 75.21±1.14 & 53.44±1.32 \\
    CPC   & 86.43±1.41 & 83.27±1.66 & 96.61±0.43 & 94.44±0.76 & 83.10±1.22 & 73.31±0.73 & 69.11±3.30 & 50.22±1.78 \\
    TNC   & 89.12±0.81 & 88.67±0.49 & 95.44±0.21 & 95.21±0.43 & 82.97±0.94 & 71.34±0.91 & 78.01±2.42 & 60.32±1.93 \\
    TS-TCC & 91.89±0.89 & 89.91±0.44 & 97.65±0.41 & 95.74±0.29 & 83.31±1.36 & 72.47±0.46 & 76.33±2.20 & 62.21±1.19 \\
      TS2Vec & 90.44±0.87 & 88.42±0.24 & 97.67±0.32 & 96.01±0.47 & 83.07±1.49 &  71.29±0.73 & 78.41±1.92 & 63.09±1.44 \\
    \midrule
    TRLS (ours)  & \textbf{93.61±0.73} & \textbf{91.23±0.27} & \textbf{97.92±0.22} & \textbf{96.02±0.31} & 85.40±0.82 & \textbf{76.76±0.41} & \textbf{88.73±1.51} & \textbf{68.83±0.32} \\

    \bottomrule
    \end{tabular}%

  \label{tab:addlabel}%
\end{table*}%

\begin{figure}[htbp]
\setlength{\belowcaptionskip}{-0.4cm}
\centering
\includegraphics[width=.5\textwidth]{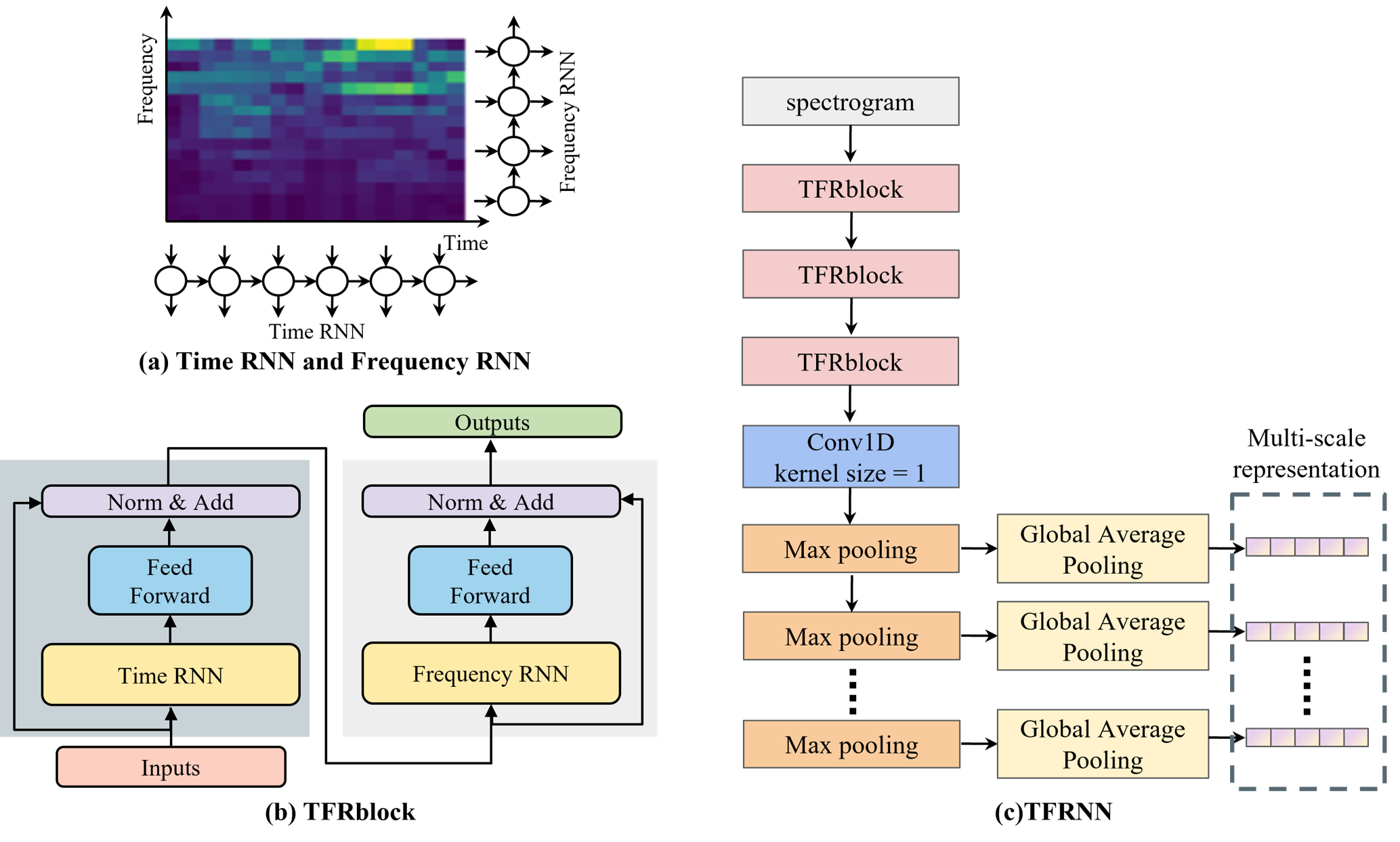}
\caption{Architecture of TFRNN model.}
\label{fig:traditional}
\end{figure}

\section{Experimental Setup}
\subsection{Datasets}
To verify the effectiveness of our framework, we compare our framework with the current optimal time series representation learning framework on four public datasets. Specifically, they are the UCI HAR dataset, the Epileptic Seizure Recognition dataset, the Sleep-EDF dataset and the ECG Waveform.

\noindent
\textbf{UCI HAR dataset} \cite{anguita2013public} collects 6 different behavior states of 30 people, including walking, walking stairs, downstairs, standing, sitting, and lying down. In addition, the sampling rate of these time series is 50Hz.

\noindent
\textbf{Epileptic Seizure Recognition dataset} \cite{andrzejak2001indications} contains EEG signals of 500 subjects, with each EEG signal lasting approximately 23.6 seconds. The dataset originally consisted of five classifications, but only one of these pertained to epilepsy. Therefore, similar to previous work \cite{eldele2021time}, we perform the binary classification task to predict epilepsy classification for this dataset.

\noindent
For the dataset of \textbf{Sleep-EDF} \cite{goldberger2000physiobank}, we mainly predict the human Sleep Stage. The data type of this dataset is the EEG signal. It has five classes: Wake (W), Non-rapid eye movement (N1, N2, N3), and Rapid Eye Movement (REM).

\noindent
\textbf{ECG Waveform} \cite{moody1983new} is a real-world clinical dataset from MIT-BIH Atrial Fibrillation, it includes 25 long-term Electrocardiogram (ECG) recordings (10 hours in duration) of human subjects with atrial fibrillation. This dataset has 4 classes (i.e. Atrial fibrillation, Atrial flutter, AV junctional rhythm, and all other rhythms) and it includes two ECG signals with 250Hz sampling rate. As the previous work, we collect one piece of data every 2500 sampling points. In addition, the data distribution is unbalanced. The AV functional rhyme is less than 0.1$\%$ which will bring many challenges to downstream tasks. Thus this dataset is very helpful for us to further study the impact of unbalanced data distribution on contrastive learning.

\subsection{Implementation Details}
We divide the four datasets into the training set, valid set and test set according to 6:2:2 ratio. For all experiments, we conduct 5-fold cross-validation, and report the mean and standard deviation. In the pre-training stage, the epoch is set to 50 and the batch size is 32. The epoch and batch size are 100 and 128 in the downstream tasks. We use Adam optimizer with a learning rate of 3e-4, weight decay of 1e-4, $\beta$1 = 0.9, and $\beta$2 = 0.99. We design two different augmentations. Particularly, our target network’s moving average decay $\tau$ = 0.7. We also set RNN's dropout to 0.2 and the number of multi-scale representations K to 5. When STFT is performed on input data, the window size of FFT for short time series sets (HAR, Epilepsy) is 32, while that for long time series sets (Sleep-EDF, ECG Waveform) is 128. Moreover, we build our framework using PyTorch 1.10 and train it on a NVIDIA GeForce RTX 2080 Ti GPU.

\section{Results}

Section 5.1 presents the evaluation of our proposed TRLS framework in comparison with five state-of-the-art time series representation learning methods using linear evaluation, as well as the effect of finetuning TRLS with less data. The performance evaluation is based on two metrics, accuracy (ACC) and macro-averaged F1-score (MF1), which provides a better indication of the model's robustness. In Section 5.2, we assess the effectiveness of TFRblock in the representation learning frameworks with time-domain time series as input. In Section 5.3, we conduct ablation experiments and investigate the impact of different data augmentation methods used in TRLS. Finally, we showcase the superior robustness of TRLS by means of representations visualization and a series of quantitative experiments, which in turn indicates that TRLS generates better representations.

\begin{table*}
  \centering
  
    \begin{tabular}{ccccccc}
    \toprule
          & Framework & Encoder & HAR   & Epilepsy & Sleep-EDF & ECG Waveform \\
        \toprule
    \multirow{4}[4]{*}{Supervised} & \multirow{2}[2]{*}{TS-TCC} & TFR   & \textbf{91.83±1.23} & \textbf{97.65±0.35} & \textbf{83.7±0.27} & \textbf{ 83.21±0.44} \\
          &       & TCN   & 90.14±2.49 & 96.66±0.24 & 83.41±1.44 &  82.43±1.32 \\
\cmidrule{2-7}          & \multirow{2}[2]{*}{TNC} & TFR   & \textbf{92.52±1.33} & \textbf{96.37±1.22} & \textbf{85.2±1.32} & \textbf{ 85.33±0.27} \\
          &       & TCN/RNN & 92.03±2.48 & 94.81±0.28 & 83.72±0.74 &  84.81±0.28 \\
    \midrule
    \multicolumn{1}{c}{\multirow{4}[4]{*}{Linear evaluation}} & \multirow{2}[2]{*}{TS-TCC} & TFR   & \textbf{91.89±0.89} & \textbf{97.65±0.41} & \textbf{83.31±1.36} & \textbf{76.33±2.20} \\
          &       & TCN   & 90.37±0.34 & 97.23±0.10 & 83.00±0.71 &  74.81±1.10 \\
\cmidrule{2-7}          & \multirow{2}[2]{*}{TNC} & TFR   & \textbf{89.12±0.81} & \textbf{95.44±0.21} & \textbf{82.97±0.94} & \textbf{78.01±2.42} \\
          &       & TCN/RNN & 88.32±0.12 & 93.22±0.42 & 81.33±0.33 & 77.79±0.84 \\
    \bottomrule
    \end{tabular}%
    \caption{The results of different encoders in the TS-TCC or TNC with supervised and linear classifier evaluation.}
  \label{tab:addlabel}%
\end{table*}%
\begin{table*}[t]
  \centering
  
    \begin{tabular}{ccccc}
    \toprule
    \multirow{2}[4]{*}{} & \multicolumn{2}{c|}{Sleep-EDF} & \multicolumn{2}{c}{ECG Waveform} \\
\cmidrule{2-5}          & ACC   & \multicolumn{1}{c|}{MF1} & ACC   & MF1 \\
    \midrule
    \textbf{TRLS} & \textbf{85.40±0.82} & \textbf{76.76±0.41} & \textbf{88.73±1.51} & \textbf{68.83±0.32} \\
    w/o spetcrogram  & 83.43±0.85 & 73.11±0.53 & 79.11±1.13 & 62.33±0.82 \\
    w/o mutli-scale representations & 83.80±1.72 & 74.79±0.41 & 85.29±2.57 & 63.94±1.75 \\
    w/o ColorJitter & 84.70±0.46 & 75.44±0.64 & 86.20±1.27 & 66.91±0.83 \\
    w/o RandomHorizontalFlip & 83.93±0.64 & 74.96±0.57 & 85.61±1.39 & 64.23±0.96 \\
    w/o GaussianBlur  & 84.92±1.44 & 75.82±0.55 & 86.54±0.73 & 67.18±0.74 \\
    w/ Negative samples  & 84.77±0.91 & 75.21±0.59 & 87.93±0.82 & 66.92±0.99 \\
    \midrule
    \multicolumn{5}{c}{\textit{Encoder}} \\
    \midrule
    \textbf{LSTM} &       &       &       &  \\
    → GRU & 85.21±0.87 & 76.03±0.62 & 87.47±0.89 & 67.73±0.51 \\
    → RNN & 84.03±0.99 & 76.76±0.41 & 86.09±0.71 & 65.21±0.85 \\
    → TCN & 83.01±1.41 & 72.93±1.31 & 82.12±2.33 & 61.31±1.71 \\
    \midrule
    \multicolumn{5}{c}{\textit{Augmentation}} \\
    \midrule
    \textbf{Different spetcrogram augmentation} & \multicolumn{4}{c}{} \\
    Same spetcrogram augmentation & 84.32±1.47 & 75.47±0.93 & 87.63±1.22 & 67.46±0.51 \\
    Different time domain augmentation (jitter\&scaling , permutation) & 79.71±0.62 & 72.93±1.21 & 79.94±1.46 & 62.03±2.12 \\
    Same time domain augmentation(jitter\&scaling) & 77.39±2.02 & 66.41±2.43 & 79.64±3.11 & 60.82±1.97 \\
    Same time domain augmentation( permutation ) & 72.62±3.11 & 63.18±2.93 & 71.87±2.45 & 56.33±3.11 \\
    Time domain augmentation \& spetcrogram augmentation & 81.66±0.98 & 72.93±1.21 & 83.61±1.49 & 63.94±1.09 \\
    \bottomrule
    \end{tabular}%
    \caption{Ablation study of each component in TRLS and data augmenation performed with linear classifier evaluation experiment}
    
  \label{tab:addlabel}%
\end{table*}%

\subsection{Comparisons with Baseline Approaches}
In our study, we compare our proposed time series representation learning framework, TRLS, with five state-of-the-art frameworks (SRL, CPC, TNC, TS-TCC, and TS2Vec) \cite{ching2018opportunities,franceschi2019unsupervised,tonekaboni2021unsupervised,eldele2021time,yue2021ts} and the supervised method. To ensure fairness in the comparison, we use the similar encoder architecture to TFRNN for the five frameworks, with only TCN added to the first layer, and kept the parameters of the two encoders (TRLS and other frameworks) consistent.  We evaluate the performance of our framework using the standard linear benchmarking evaluation scheme in Table 1. Our results show that TRLS achieve the best performance in HAR, Epilepsy, and ECG Waveform datasets. For Sleep-EDF, although our accuracy evaluation index is slightly lower than the supervised method, our MF1 evaluation index is higher, indicating that our framework generated representations with stronger generalization ability. Additionally, in the five-fold cross-validation experiments, TRLS exhibits the smallest variance, indicating that the representations generated by TRLS are robust enough to maintain stable performance across different training data. We also find that although five state-of-the-art frameworks have the excellent results on shorter time series datasets (HAR, Epilepsy), approaching or even exceeding supervised method. However, there is a large decline in processing datasets with long time series (Sleep-EDF, ECG Waveform). This is due to the sparse characteristic points of long time series in the time domain, which makes it challenging to extract effective representations. In contrast, our spectrogram-based data input method aggregates information from both the time and frequency domains, resulting in more distinctive data. As a result, the encoder can easily extract more differentiated representations.

\begin{figure}[htb]
\setlength{\belowcaptionskip}{-0.4cm}
\centering
\includegraphics[width=.48\textwidth]{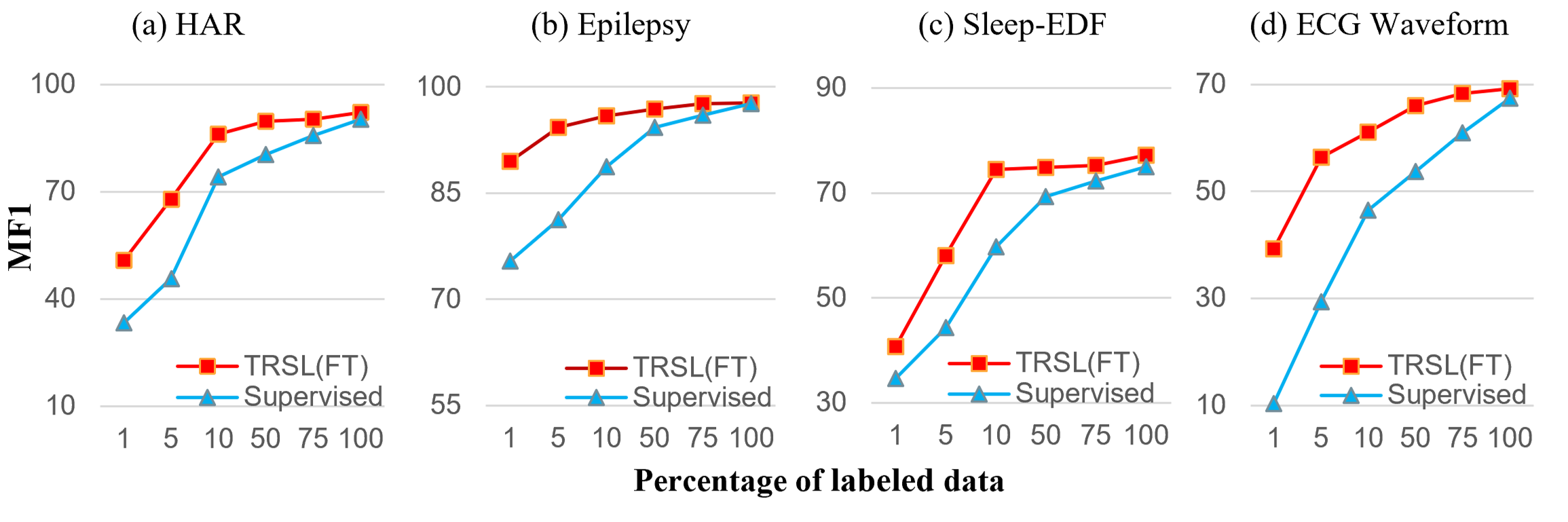}
\caption{Comparison between supervised training vs TRLS fine-tuning for different few-labeled data scenarios in terms of MF1.}
\label{fig:traditional}
\end{figure}

For investigating the effectiveness of our TRLS under semi-supervised settings, we train the model with 1$\%$, 5$\%$,10$\%$, 50$\%$, and 75$\%$ of randomly selected instances from the training data. Figure 5 depicts the results of TRLS and the supervised training under these settings. In particular, TRLS finetune (i.e., red curves in Figure 5) means that a few labeled samples are employed to finetune the pre-trained encoder. We observe that the performance of supervised training declines seriously under the limited data. However, with only 10$\%$ data by TRLS fine-tuning, we can achieve supervised training performance close to 100$\%$ with the four datasets. This demonstrates the effectiveness of TRLS under semi-supervised settings.


\subsection{The effectiveness of TFRblock}
In this section, we evaluate the performance of TFRblock in the existing time series representation learning framework TS-TCC and TNC’s downstream classification tasks. We modify the encoders of these two frameworks. In the encoder of TS-TCC (TCN), we replaced all the TCN layers except the first one with TFRblock. Note that, the parameters of modified encoder are consistent with those of the original one in TS-TCC. For TNC, which employs TCN and RNN encoders for different datasets, we use a similar encoder modified from TS-TCC for comparison, and keep the same parameter quantity as the original encoder of TNC.

The results of our experiments on four datasets are presented in Table 2. Our TFRblock outperforms the original TCN and RNN encoders in both supervised and linear evaluation tasks in TS-TCC and TNC frameworks, respectively. This implies that our TRFblock is a better feature extraction module. It can achieve good performance not only on spectrogram data but also on time domain data as input in contrastive learning frameworks. Furthermore, the use of TFRblock narrows the performance gap between the supervised and linear evaluation tasks under both frameworks, indicating that the representations extracted by TFRblock are robust.

\subsection{Ablation Study}
To verify the effectiveness of the proposed components in TRLS, a comparison between TRLS and its six components on Sleep-EDF and ECG Waveform is given in Table 3. The six components are as follows: (1) w/o spectrogram: the input data is changed from spectrogram to time domain data. Besides, the data augmentation method is the same as TS-TCC, and the encoder is the modified encoder in Section 5.1. (2) w/o multi-scale representations change multi-scale representations to single representation (3) w/o ColorJitter, w/o RandomHorizontalFlip, and w/o Gaussian Blur remove the corresponding data augmentations respectively. (4) w/ Negative samples: negative samples are introduced in the TRLS, and loss function is the same as that used in TS-TCC. As shown in table 3, first, the performance of TRLS in time domain data is significantly reduced, which indicates the advantage of using spectrogram as input. Second, our framework is highly sensitive to multi-scale representations. The learning of multi-scale representations is a form of data augmentation that provides more representations for training the encoder, it makes the learned encoder more robust. Third, RandomHorizontalFlip has a greater impact on our framework's performance than the other two data augmentations. Fourth, the introduction of negative samples leads to the degradation of the TRLS-trained encoder's performance, demonstrating that improper negative samples design reduces the robustness of the generated representations. Moreover, table 4 displays the effect of modifying LSTM in TFRblock to GRU, RNN and TCN respectively. The results show that both RNN and TCN significantly decrease compared to LSTM, while GRU decreases only slightly.

We conduct different types data augmentation experiments on the spectrogram and time domain augmentations. The spectrogram augmentation directly applies data augmentation to the spectrogram, while the time domain augmentation involves transforming the time domain data into a spectrogram after time domain data augmentation. We also perform six comparative tests. (1) Different spectrogram augmentations are the data augmentations designed in Section 4.2. (2) Same spectrogram augmentations represents that the input data uses a random variant of same spectrogram augmentations. The data augmentations used here are ColorJitter, RandomHorizontalFlip, and Gaussian Blur. (3) Different time domain augmentations adopts the same time domain augmentation as \cite{eldele2021time}. (4), (5) Same time domain augmentation means that only one kind of time domain augmentation is made for the input data. (6) Time domain augmentation $\&$ spectrogram augmentation means that we use time domain augmentation and spectrogram augmentation for input data together. Our results show that the performance of spectrogram augmentation is better than that of time domain augmentation. Moreover, the method of time domain augmentation $\&$ spectrogram augmentation used together leads to performance degradation in our scheme.


\subsection{Representations Visualization and Quantitative Experiments}

We use the TSNE map to visualize the representations generated by various frameworks (TNC, TS-TCC, TS2Vec, TRLS). As depicted in the Figure 6, it is clear that the representations generated by our proposed TRLS framework possess superior discriminability compared to the other frameworks. This observation is highly beneficial for downstream tasks, as it suggests that the learned representations are better suited for classification and other related tasks.

\begin{figure}[htb]
\setlength{\belowcaptionskip}{-0.4cm}
\centering
\includegraphics[width=.48\textwidth]{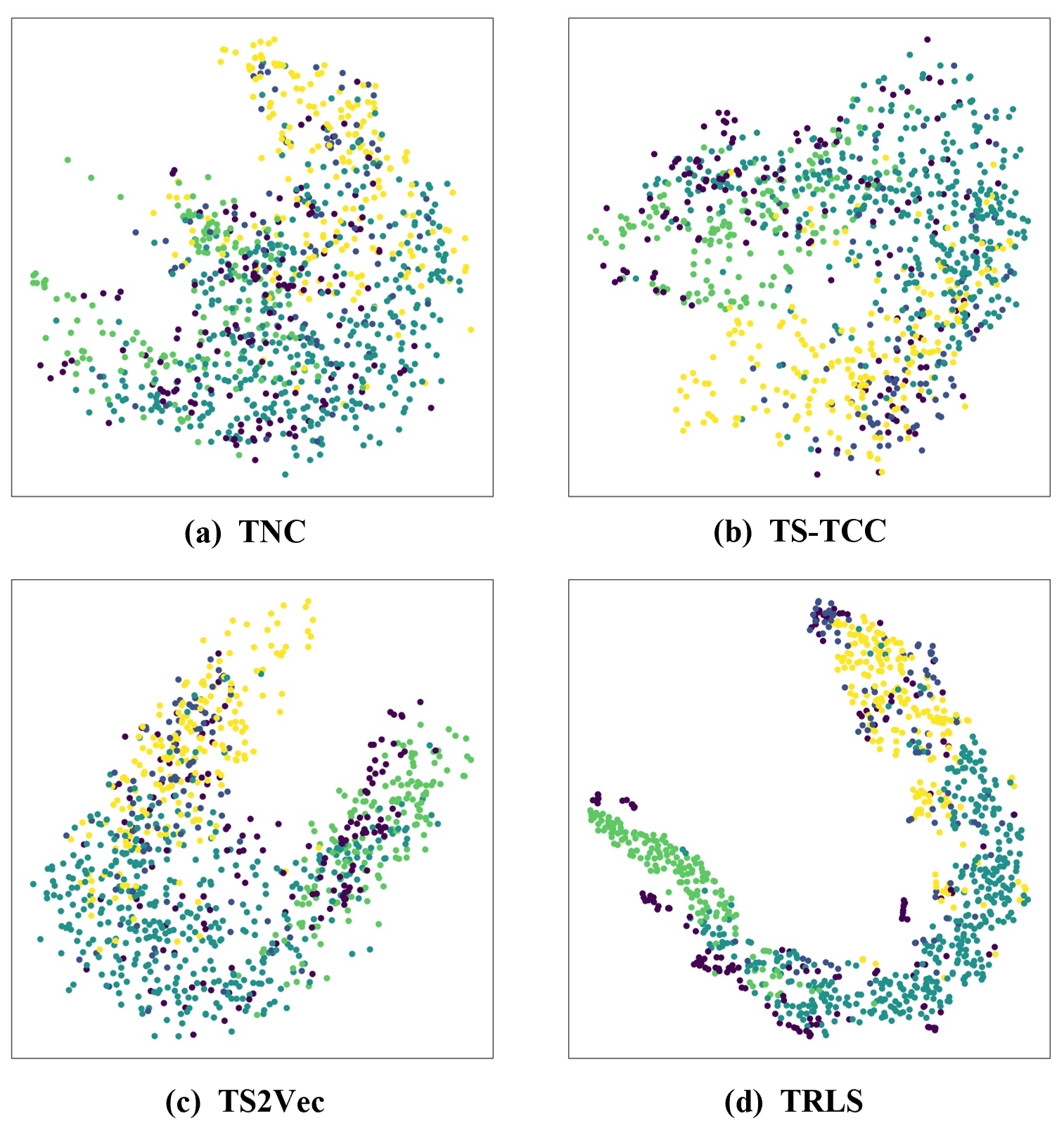}
\caption{TSNE map of the time series representations from different frameworks in the Sleep-EDF.}
\label{fig:traditional}
\end{figure}

We perform a series of quantitative experiments on TRLS to show the robustness of it. Firstly, we verify the impact of the representation augmentation method on the robustness of the encoder.  We adjust K that is the number of multi-scale representations and then observe the results in the ECG Waveform dataset. With the increase of K, the MF1 value increases, which indicates that our multi-scale representations effectively expand the space of representation and significantly improve the generalization of encoder in unbalanced datasets. However, accuracy decreases when K $\textgreater$ 5. To ensure the balance between accuracy and MF1, our framework chooses K = 5.

\begin{table}
  \centering
  
    \begin{tabular}{ccc}
    \toprule
    \multirow{2}[4]{*}{K} & \multicolumn{2}{c}{ECG Waveform} \\
\cmidrule{2-3}          & ACC   & MF1 \\
    \midrule
    1     & 85.29±2.57 & 63.94±1.75 \\
    3     & 86.97±0.74 & 66.61±0.87 \\
    5     & \textbf{88.73±1.51} & 68.83±0.32 \\
    7     & 87.92±1.62 & 69.07±0.49 \\
    9     & 88.02±1.57 & 69.14±0.26 \\
    11    & 87.44±1.88 & \textbf{69.19±0.33} \\
    \bottomrule
    \end{tabular}%
    \caption{Effect of K on robustness.}
  \label{tab:addlabel}%
\end{table}%
In the end, we evaluate TRLS's performance on Sleep-EDF dataset by applying dropout and adding Gaussian noise with different Signal to Noise Ratio (SNR). The linear evaluation results are presented in Table 5, indicating that the performance of TRLS does not decrease significantly when the dropout rate is less than or equal to 0.6. Moreover, we add Gaussian noise with varying SNR to the data, and the results showed that TRLS maintains the accuracy above 81$\%$ even under extreme SNR conditions (e.g., SNR=0.01). These experimental results demonstrate that TRLS exhibits good robustness to data sparsity and noise.

\begin{table}
  \centering
  \setlength{\tabcolsep}{1mm}{
  \scalebox{1.0}{
    \begin{tabular}{cccccc}
    \toprule
    \multirow{2}[4]{*}{dropout} & \multicolumn{2}{c}{Sleep-EDF} & \multirow{2}[4]{*}{SNR(db)} & \multicolumn{2}{c}{Sleep-EDF} \\
\cmidrule{2-3}\cmidrule{5-6}          & ACC   & MF1   &       & ACC   & MF1 \\
    \midrule
    0     & \textbf{85.40±0.82} & \textbf{76.76±0.41} & -     & \textbf{85.40±0.82} & \textbf{76.76±0.41} \\
    0.1   & 85.14±0.62 & 75.3±0.42 & 10    & 84.21±0.84 & 75..01±1.54 \\
    0.2   & 85.11±0.75 & 74.6±0.54 & 5     & 83.39±1.11 & 73.65±0.37 \\
    0.3   & 84.52±0.96 & 74.24±0.98 & 1     & 82.02±1.74 & 72.01±1.33 \\
    0.4   & 83.61±1.08 & 73.58±0.46 & 0.9   & 81.20±0.38 & 71.12±0.32 \\
    0.5   & 82.87±0.83 & 72.36±0.58 & 0.7   & 81.85±0.43 & 71.67±0.55 \\
    0.6   & 81.19±1.33 & 71.12±0.97 & 0.5   & 81.71±0.19 & 71.53±0.83 \\
    0.7   & 79.27±0.70 & 70.02±1.41 & 0.3   & 81.54±0.38 & 71.37±0.62 \\
    0.8   & 76.53±1.24 & 66.41±0.50 & 0.1   & 81.30±0.41 & 71.22±0.61 \\
    0.9   & 66.53±2.51 & 58.44±2.06 & 0.01  & 81.26±0.77 & 71.08±0.07 \\
    \bottomrule
    \end{tabular}}}%
        \caption{TRLS performance in time series with different sparsity and SNR.}
  \label{tab:addlabel}%
\end{table}%

\section{Conclusion}
This paper proposes a novel framework called TRLS for better learning appropriate representations. TRLS focuses on problems that are not concerned in the current mainstream frameworks: complexity and sparsity of time series, more robust encoders and  the problem of constructing negative samples. The spectrogram is firstly introduced into our framework to alleviate problem of sparsity and complexity of time series. The spectrogram is transform into two different views by two different data augmentations. Then the encoder is designed to generates multi-scale representations from the augmented spectrogram. In this way, the robustness of encoder is enhanced. Next, multi-scale representations for subsequent contrastive learning are used and two views are flipped to calculate loss again. Our framework can guide encoder training without negative samples. Thus the TRLS avoids the problems caused by designing inappropriate negative samples. Finally, a series of experiments are conducted on four public datasets, and the results show that our framework is prior to all the existing ones on all the used datasets.

\end{document}